\documentclass{article}

\input{tcilatex}

\begin{document}

\title{{\LARGE Strong Phases in the Decays }$B${\LARGE \ to }$D\pi $}
\author{{\large Lincoln Wolfenstein} \\
{\small Department of Physics, Carnegie Mellon University, Pittsburgh, PA
15213}}
\maketitle

\begin{abstract}
The observed strong phase difference of 30$^{\text{o}}$ between $I=\frac{3}{2%
}$ and $I=\frac{1}{2}$ final states for the decay $B\rightarrow D\pi $ is
analyzed in terms of rescattering like $D^{\ast }\pi \rightarrow D\pi ,$
etc. \ It is concluded that for the decay $B^{o}\rightarrow D^{+}\pi ^{-}$
the strong phase is only about 10$^{\text{o}}$. \ Implications for the
determination of $\sin \left( 2\beta +\gamma \right) $ are discussed.
\end{abstract}

\qquad \vspace{1in}

The weak decay amplitude to a specific final state can be written as $%
Ae^{i\delta }$ where $A$ is the decay amplitude, in general complex, and $%
\delta $ is the "strong phase". \ For a final eigenstate $\delta $ is simply
the elastic scattering phase in accordance with the Watson theorem. \ For
the case of $B$ decays the final scattering is primarily inelastic; $\delta $
arises from the absorptive part of the \ decay amplitude corresponding to a
weak decay to intermediate states followed by a strong scattering to the
final state. \ Many papers have discussed the expected size of $\delta $ 
\cite{Beneke}.

\qquad Recently data has suggested a significant non-zero phase for the
decays $\overline{B}$ to $D\pi $ \cite{Ahmed}. \ These decays can be
analyzed in terms of two amplitudes $A_{3}\ e^{i\delta _{3}}$ and $A_{1}\
e^{i\delta _{1}}$ corresponding to the final isospin states $\dfrac{3}{2}$
and $\dfrac{1}{2}.$ \ The amplitudes for the three decays of interest are
the following:

\begin{eqnarray}
A_{o-} &=&A\left( D^{\text{o}}\ \pi ^{-}\right) =A_{3}e^{i\delta _{3}} 
\TCItag*{(1a)} \\
A_{+-} &=&A\left( D^{+}\ \pi ^{-}\right) =\frac{1}{3}A_{3}\ e^{i\delta _{3}}+%
\frac{2}{3}A_{1}\ e^{i\delta _{1}}  \TCItag*{(1b)} \\
A_{oo} &=&A\left( D^{\text{o}}\pi ^{\text{o}}\right) =\frac{\sqrt{2}}{3}%
\left( A_{3}e^{i\delta _{3}}-A_{1}\ e^{i\delta _{1}}\right)  \TCItag*{(1c)}
\end{eqnarray}

The experimental result gives the ratio of decay probabilities

\[
D^{\text{o}}\pi ^{-}\colon D^{+}\pi ^{-}\colon D^{\text{o}}\pi ^{\text{o}%
}=46\colon 27\colon 2.9 
\]

From this one deduces

\begin{equation}
\frac{A_{1}}{A_{3}}=0.69,\quad \cos \left( \delta _{3}-\delta _{1}\right)
=0.86  \tag*{(2)}
\end{equation}

Thus we find approximately a 30 degree phase difference, significantly
different from zero. \ We discuss here the implications of such a phase
difference; there remain, of course, sizable errors on this value.

We now make the assumption that the major rescattering comes from states of
the form $D_{i}^{+}\pi _{j}^{-}$ such as $D^{\ast +}\pi ^{-},\ D^{+}\rho
^{-},\ $etc.\cite{Chuva}\ \cite{Falk} \ Such states are expected in
factorization and about 10\% of the $b$ to $u\bar{c}\ d$ \ transitions have
been identified to be of this type. \ It is much less likely that
complicated many-particle states should rescatter to $D\pi .$

We consider first the simple factorization (large $N_{c}$) limit:

\[
A_{+-}=A_{o-}\qquad A_{oo}=0 
\]

Here the $\pi ^{-}$ is assumed to come directly from the $\bar{u}\ d$
current, the amplitudes are real and there is no $D^{\text{o}}\pi ^{\text{o}%
} $ decay. \ This corresponds to $A_{3}=A_{1}=A.$ \ We now add rescattering
from states of the form $D_{i}^{+}\pi _{j}^{-}.$ \ We label these amplitudes 
$X_{3}\ $and $X_{1}$ and again $X_{3}=X_{1.}$ The most obvious rescattering
occurs via the exchange of an isospin 1 particle, either $\pi $ or $\rho .$
\ As a result the rescattering amplitude is proportional to $\tau _{i}\cdot
T_{j}$ with the values $\left( \frac{1}{2},\ -1\right) $ for $I=\left( \frac{%
3}{2},\frac{1}{2}\right) .$ \ The resultant imaginary amplitudes

\begin{equation}
\frac{\func{Im}\ A_{3}}{\func{Im}\ A_{1}}=\frac{X_{3}\ \left( \frac{1}{2}%
\right) }{X_{1}\left( -1\right) }=-\frac{1}{2}  \tag*{(3)}
\end{equation}

Considering the $\func{Im}\ A_{i}$ as fairly small, this means

\begin{equation}
\delta _{1}=-2\delta _{3}  \tag*{(4a)}
\end{equation}%
If we use the empirical value $\left( \delta _{3}-\delta _{1}\right) =30^{%
\text{o}}$

\begin{equation}
\delta _{3}=10^{\text{o}}\qquad \delta _{1}=-20^{\text{o}}  \tag*{(4b)}
\end{equation}%
and to lowest order in $\delta _{1}$ and $\delta _{3}$

\begin{eqnarray*}
A_{o-} &=&Ae^{i\delta _{3}}\quad ,\qquad A_{+-}=Ae^{-i\delta _{3}} \\
A_{oo} &=&\sqrt{2}\quad iA\ \delta _{3}
\end{eqnarray*}

Thus the phase difference of 30$^{\text{o}}$ corresponds to a fairly small
phase of magnitude 10$^{\text{o}}$ for both of the favored decays. \ Of
course in this approximation the $D^{\text{o}}\pi ^{\text{o}}$ decay is
purely imaginary entirely due to rescattering.

If we now use the empirical value $A_{1}=0.7A_{3}=0.7A$ \ but still assume $%
X_{3}=X_{1},$ we have using Eq. (1) to lowest order in $\delta _{1}$ and $%
\delta _{3x}.$

\begin{eqnarray*}
A_{3} &=&A+i\ \delta _{3}A \\
A_{1} &=&0.7A-2i\ \delta _{3}A \\
\delta _{1} &=&\frac{-2\delta _{3}}{0.7} \\
A_{+-} &=&0.8A-iA\ \delta _{3}
\end{eqnarray*}%
and with $\delta _{3}-\delta _{1}=30^{\text{o}}$ we get $\delta _{1}=-22^{%
\text{o}}\quad ,\quad \delta _{3}=8^{\text{o}}$ \ \ and the phase for $\bar{B%
}\rightarrow D^{+}\pi ^{-}\ $is again $10^{\text{o}}.\bigskip $ \ 

Finally if we also assume$\ X_{1}=0.7\ X_{3}$ we get Eqs. (4) again and

\[
A_{+-}=0.8A-i0.6A\ \delta _{3} 
\]%
and the phase for $\bar{B}\rightarrow D^{+}\pi ^{-}\ $is $7.5^{\text{o}}.$

\qquad Thus we conclude that $\left( \delta _{3}-\delta _{1}\right) =30^{%
\text{o}}$ corresponds to a small phase of order $10^{\text{o}}$ for $\bar{B}%
\rightarrow D^{+}\pi ^{-}.$ \ In contrast the phase for the unfavored decay $%
\bar{B}\rightarrow D^{\text{o}}\pi ^{\text{o}}$ is greater than $45^{\text{o}%
}.$ \ Similar results are implied by more detailed analysis \cite{Chuva}.

\qquad One reason for the interest in the strong phase for $\bar{B}%
\rightarrow D^{+}\pi ^{-}$ is the possible use of this decay or the related
decay $\bar{B}\rightarrow D^{\ast +}\ \pi ^{-}$ in the determination of the
phase $\gamma $ in the CKM matrix. \ One can look at the time-dependence of
the decay due to interference with the double-Cabibbo suppressed decay $%
B\rightarrow D^{+}\pi ^{-}$ which corresponds to $\bar{b}\rightarrow \bar{u}%
+c+\bar{d}.$ \ The time-dependent term can be used \cite{Dunietz} to find $%
\sin \left( 2\beta +\gamma \right) .$ \ The detailed analysis \cite{Suprun}\ 
\cite{Silva} involves the strong phase $\Delta ,$ which is the difference
between the strong phase for $\bar{B}\rightarrow D^{+}\pi ^{-}$ and that for 
$B\rightarrow D^{+}\pi ^{-}.$ \ There is an ambiguity in the result unless
one can assume $\Delta $ is small.

The same isospin analysis given for $\bar{B}\rightarrow D^{+}\pi ^{-}$ can
be applied to $B\rightarrow D^{+}\pi ^{-}$ and one expects again that the
final state phase is due to the same rescattering from status like $D^{\ast
}\pi ,\ D\rho ,\ $etc. \ The relative importance will be different for the
case of $B$ as compared to $\bar{B},$ but theoretical estimates \cite{Suprun}%
\ indicate the difference is not large. \ Thus $\Delta $ is expected to
innvolve a cancellation between the two strong phases and thus be smaller
than either one. \ Given our conclusion that the phase for $\bar{B}%
\rightarrow D^{+}\pi ^{-}$ is of order $10^{\text{o}}$ \ we conclude that $%
\Delta $ is very small.

All the analysis here holds equally well for the decays to $D^{\ast }\pi .$
\ In fact the experimental results \cite{Coan} for the decays to $D^{\ast
}\pi $ are the same within errors as for $D\pi $ and give essentially the
same strong phase shift Eq. (2).

Decays in which the final $\pi ^{-}$ is replaced by a $\rho ^{-}$ are found
to have a branching ratio 2 to 3 times as large as those with a $\pi ^{-}.$
\ Thus it may be expected that rescattering from states $D_{i}\rho $ to $%
D_{i}\pi $ may have a larger effect than $D_{i}\pi $ to $D_{i}\rho .$ \ Thus
while our general analysis might be applicable to $D_{i}\rho $ we expect the
magnitude of the strong phase shifts would be smaller. \ This seems to be
true from the first data on the $D^{\text{o}}\rho ^{\text{o}}$ decay \cite%
{Sapathy}.

This research was supported in part by the Department of Energy under grant
no. DE-FG02-91ER40682. \ Part of this work was carried out at the Aspen
Center for Physics.

\end{document}